\documentclass{article}
\font\cero=cmss10 scaled 1728 
\usepackage[utf8]{inputenc} 
\usepackage{multirow}\usepackage{tensor}
\usepackage{verbatim}
\usepackage{amsfonts}
\usepackage{graphicx}
\usepackage{amssymb}
\usepackage{float}
\begin{document}
\begin{flushleft}
{\cero Hyperbolic symmetries, inflaton-phantom cosmology, and inflation}\\
\end{flushleft} 
{\sf R. Cartas-Fuentevilla, A. Escalante-Hernandez, and A. Herrera-Aguilar}\\
{\it Instituto de F\'{\i}sica, Universidad Aut\'onoma de Puebla,
Apartado postal J-48 72570, Puebla Pue., M\'exico}; 

\noindent{\sf R. Gonzalez-Cuaglia} \\
{\it Universidad de las Am\'ericas Puebla, Ex-hda Sta. Catarina M\'artir, C.P.
 72810, San Andr\'es Cholula, Pue., M\'exico.} \\

ABSTRACT: Using a hyperbolic complex plane, we study the realization of the underlying hyperbolic symmetry as an internal symmetry that enables the unification of scalar fields of cosmological and particle physics interest. Such an unification is achieved along the universal prescriptions used in physics, avoiding the use of concepts as Euclideanization, non-canonical Lagrangians and hidden structures, that have appeared in other approaches. The scalar potentials constructed within the present scheme are bounded from below, and the realization of the spontaneous symmetry breaking of the aforementioned noncompact symmetry is studied. The profiles of these potentials with exact/broken hyperbolic symmetry replicate qualitative aspects of those ones used in inflationary models, and then a detailed com\-pa\-ri\-son is made.
Moreover, the homotopy constraints of the topology induced on the corresponding vacuum manifolds, restricts the existence of topological defects associated with continuous symmetries, allowing only those defects 
associated with discrete symmetries;  the consistency of these results is contrasted with current observational tests from the LIGO/Virgo collaboration, and terrestrial experiments based on a synchronized network of atomic magnetometers. At the end,
the nonre\-la\-tivistic limit of the model is identified with a hyperbolic version of the nonlinear Schr\"odinger equation.
\\


\section{Motivations, antecedents, and results}

In any field theory, unification plays an important role, both at the level of physical simplicity, and at the level of mathematical beauty. For years, the idea of unification has served as a guideline towards the construction of unified field theories, successfully implemented within the framework of particle physics, for instance. Within this paradigm, two formerly unrelated branches of physics are combined into a single conceptual formalism by a new theoretical perspective. This was the case for electricity and magnetism, optics and electromagnetism, thermodynamics and statistical physics, inertial and gravitational forces, among other classical examples (see \cite{goenner} for a historical review on classical aspects of unified field theories). 

Within a more fundamental quantum level, electromagnetic and weak nuclear forces were unified into the electroweak force, subsequently, the strong interaction, known as chromodynamics, was also included into the same setup leading to the standard model of elementary particle physics. Unification with gravity, the fourth fundamental interaction, is the aim of considerable current research within several branches of physics, namely, classical general relativity, supergravity, superstring and super membrane theories, leading to new and deep developments both in theoretical physics and mathematics and must be considered nowadays as highlights of physical research \cite{goenner}. 

Within this context, two interdependent approaches regarding unification can be highlighted: the first one focuses on the unification of the representations of physical fields: special relativity encompasses all phenomena which involve velocities close to the velocity of light in vacuum, providing a synthesis of the laws of mechanics and electromagnetism; quantum field theory unifies classical fields and the quantum realm. The second line of thought seeks for unification of the dynamics of physical fields, i.e. unification of the fundamental interactions: the aforementioned Maxwell's theory and the standard model of particle physics are clear examples of this kind. There exists, however, a third mathematical-physics approach which consists of making use of a hyperbolic complex plane, for instance, in order to unveil a hyperbolic internal symmetry that supports the unification of scalar fields of both particle physics and cosmological relevance\cite{Oscar}-\cite{alex}. In fact, there are many other approaches of this type, see for instance \cite{kassandrov1}-\cite{kassandrov3} for a similar treatment using  a biquaternionoc algebra for describing Lorentz invariant Maxwell and Yang-Mills equations, and \cite{kechkin} for  a study where a complex time incorporates the arrow of time of irreversible processes within quantum field theory with a symmetry breaking mechanism, for instance.

The above mentioned unification spirit is also present regarding different scalar fields that appear within cosmology, where inflationary models described with the aid of two or more scalar fields possess interesting and even unusual properties (for an interesting review on this topic see, for instance, \cite{Riotto} and references therein). Moreover, with the purpose of unifying scalar fields of cosmological interest, an inflaton-phantom field unification was proposed in \cite{Lorenz} by considering those fields as the real and imaginary components of an ordinary complex scalar field, $\phi\equiv\phi_{1}+i \phi_{2}$, where $\phi_{1}$ represents an inflaton field that drives the inflationary stage of the early Universe, while the phantom field must be identified with $\phi_{2}$ due to its unusual dynamics, which governs the accelerated expansion of the Universe.  On the other hand, phantom fields may be necessary for solving the horizon problem in scalar cosmology \cite{fermi}.
The underlying internal symmetry that protects the inflaton-phantom field unification is $SO(1,1)$. 
This unification symmetry allows one to build successful cosmological potentials and provides an interesting matter-phantom duality which relates scalar matter cosmological solutions and phantom solutions to each other.

Although a phantom field may be admissible in cosmological scenarios, such a field is quantum-mechanically unstable; assuming that the phantom field  couples  to gravity, the quantum vacuum can decay into positive-energy gravitons and negative-energy ghosts while conserving energy, and the decay rate diverges and the theory is physically untenable
\cite{cline}. An argument for facing this conflict is, for example, by considering that the phantom field is only an effective one at long distances, and thus the decay of the vacuum is slow on cosmological time scales. Furthermore, according to the constraints obtained in \cite{cline}, phantom fields can come from  a low-energy sector completely hidden from the standard model. Additionally, from the perspective of quantum deSitter cosmology, a phantom scalar field may correspond to an effective description of some quantum field theory \cite{nojiri}.
In this manner, in the formulation at hand, the phantom field is considered, such as in the reference \cite{Lorenz},  only a classical field at cosmological level, 
and acceptable as an effective model  of a quantized canonical field along the ideas developed in \cite{nojiri}.

Our treatment will be developed in parallelism with the reference  \cite{Lorenz}, establishing similarities, important differences, and criticism; hence,
even though the guideline in \cite{Lorenz} was the use of simplicity principles, a non-canonical form for the kinetic term of the field $\phi$ was proposed,
\begin{eqnarray}
{\cal L}= \partial_{\mu}\phi\partial^{\mu}\phi+\partial_{\mu}\phi^{*}\partial^{\mu}\phi^{*}= 2\left(
\partial_{\mu}\phi_{1}\partial^{\mu}\phi_{1}-\partial_{\mu}\phi_{2}\partial^{\mu}\phi_{2}\right);
\label{non}
\end{eqnarray}
where the term for the field $\phi_{2}$ has a negative sign, and can effectively be identified with a phantom field. Furthermore, 
in order to show that the functional form of ${\cal L}$ in terms of the pair $(\phi,\phi^{*})$ is the same that in terms of the pair $(\phi_{1},\phi_{2})$, the authors considered the transformation $\phi^{*}\rightarrow i\phi^{*}$, a ``Wick rotation" in field space which leaves the field $\phi$ untouched\footnote{In fact, one should expect that both fields $\phi$ and $\phi^{*}$are consistently transformed.}; hence, the minus sign is conveniently absorbed. Later on, an Euclidean {\it vectorlike} $\Phi_{E}$ array is constructed, and the Lagrangian takes a simple form,
\begin{eqnarray}
\Phi_{E}=\left( \begin{array}{c}
                  \phi \\
                 \phi^{*}
                 \end{array} \right),
\quad
{\cal L}=\partial_{\mu}\Phi_{E}^{T}\partial^{\mu}\Phi_{E}.
\label{eu}
\end{eqnarray}
Similarly in terms of the real fields the Lagrangian was rewritten as
\begin{eqnarray}
\Phi=\left( \begin{array}{c}
                  \phi_{1} \\
                 \phi_{2}
                 \end{array} \right),
\quad
{\cal L}=\partial_{\mu}\Phi^{T}\cdot \sigma_{3}\cdot\partial^{\mu}\Phi, \quad \sigma_{3}=\{1,-1\}.
\label{min}
\end{eqnarray}
In this manner the symmetry groups are identified, the Lie group $O(2,C)$ that preserves the Euclidean metric in field space for the Lagrangian
(\ref{eu}), and the isomorphic $O(1,1,C)$ group that preserves the Minkowski metric for the Lagrangian (\ref{min}); again, the ``Wick rotation" in field space establishes the equivalence between those prescriptions. Finally the special subgroup $SO(1,1)$ is chosen as the fundamental symmetry of the theory, which protects the inflaton-phantom unification.

The complexification of fields, such as that described previously in the cosmological context, is realized traditionally by using the usual complex unit $i$; in fact, from the historical point of view, complexification has been synonymous of the use of the conventional two-dimensional complex plane. However, it is little known that actually there exist three complex units, the elliptic (the conventional one), the parabolic, and the hyperbolic one \cite{kisil}; such a classification of mathematical objects into those three classes is practically universal. In \cite{kisil} Kisil explains in detail how these three complex units emerge as particular realizations of the so called M\"obius maps, which can be fully understood within the Klein's Erlangen program.
See for example \cite{ulrich} (and references there in), for an exploration of few applications in physics of alternative complex units.

One of the purposes of the present article is to show that, specifically in the case of the inflaton-phantom unification, once we have chosen the appropriate complex unit, namely the hyperbolic one,  such an unification follows naturally along the conventional prescriptions, without invoking any additional internal structure in field space: no non-canonical kinetic terms, no Wick rotation, no vector-like arrays, no Euclidean/Minkowskian internal metrics, etc., are required.
Such prescriptions are universal and very known in physics, for example they are of common use in the construction of particle physics models. Specifically  the parallelism between the case at hand and the $U(1)$ field theory is absolute, since the kinetic terms, the mass terms, interaction terms, etc. have always the canonical form. In the next section we shall outline the basic ingredients of the hyperbolic complex plane, which will lead, in section \ref{uni}, to the aforementioned unification.

In section \ref{mexican} we shall show that quartic-order terms leading to spontaneous symmetry breaking scenarios, have exactly the same functional form in terms of the complex fields, such as the usual ``$\lambda \phi^4$" term of the $U(1)$ field theory, and can lead to potentials bounded from below, in contrast to the traditional belief of ``unbounded potentials" due to the unusual dynamics of the phantom field (see for example the unbounded potential considered in Eq. (14) of \cite{Lorenz}). Additionally the hyperbolic symmetry present in the formulation at hand will induce a change in the topology of the vacuum manifold, introducing a hyperbola instead of the circle for the $U(1)$ field theory. 

It is currently believed that continuous global symmetries are approximate (see \cite{dine} and references cited therein); explicit breaking of global symmetries can arise for example from instanton induced effects;
additionally gravity may break global symmetries through black holes physics \cite{rai}. A pragmatic way to explicitly break a global symmetry is by adding a term depending on the phase of the complex field $\phi$, {\it i.e.} a term of the form $\phi^{d}+\phi^{*d}$, where $d$ is the dimension of the background space-time. In sections \ref{break} and \ref{qua} we shall show that when the new hyperbolic symmetry is broken by means of this mechanism, 
the deformed potential is again bounded from below, and the hyperbola representing the original degenerate vacuum will undergo a {\it retraction} to two points, which represent the new vacuum,
without changing the scenario for the possible formation of certain topological defects.  Hence, one obtains a residual discrete symmetry, and a Goldstone-like boson field acquires mass, such as the pion in QCD.
Additionally the potentials with exact/broken hyperbolic symmetries obtained within the present approach, will have a desired profile for hybrid inflationary scenarios, a set of critical bifurcation points at the {\it ridge}, and points representing true minimum states, enabling in this way the possibility of the preheating phenomenon. A qualitative comparison with specific inflationary models is made in our concluding remarks.

The change of topology will have a dramatic effect on the formation of topological defects; in particular the qualitative aspects of the possible formation of topological strings and domain walls are discussed within the present formulation in section \ref{topo}, taking into account current observational evidences (in favor and against).
In section \ref{hamilton}, we develop the nonrelativistic limit of the action
with hyperbolic symmetry, and a hyperbolic version of the non-linear quantum mechanics is obtained.
In our concluding remarks we describe certain commutative rings that are endowed with algebraic structures that allow the unification of arbitrary spin fields, involving both compact and noncompact symmetries.

\section{Hyperbolic numbers}
\label{hn}
The hyperbolic numbers ${\cal H}$ correspond to a two-dimensional number system over the reals,
\begin{eqnarray}
z=x+jy, \quad \overline{j}=-j, \quad j^{2}=1; \quad
\overline{z}=x-jy;
\label{ring}
\end{eqnarray}
thus, the only difference with respect of conventional complexes, is the 
fact that the square of the hyperbolic unit $j$ is one; this property however has a profound effect on the structure of the complex plane. Furthermore, 
this new property will play a key role in the desired inflaton-phantom unification, and will allow us to reveal in a direct way the underlying symmetry of the theory. Considering the perplex conjugate $\overline{z}$, the modulus is defined as usual,
\begin{eqnarray}
z\overline{z}=x^2-y^{2};
\label{modulus}
\end{eqnarray}
where the minus sign emerges naturally due to the definition of the hyperbolic unit.
As expected, the hyperbolic numbers are closed respect to the addition, and multiplication. However, division is striking due to the presence of zero divisors: Any hyperbolic number with $x=\pm y$ is a zero divisor since its modulus vanishes, and in general
${\cal H}$ corresponds to a (commutative) ring; thus $1/z$ exists provided that $x\neq \pm y$. 

In the case of the conventional complex plane, the unit circle is defined as the set of numbers with modulus $1$, which corresponds to a non- simply connected manifold, and thus with a $\pi_{1}$ nontrivial homotopy group. In the case at hand the unit hyperbola is defined as $x^{2}-y^{2}=1$, and corresponds actually to a pair of hyperbolae, and thus a manifold formed by four branches;  each branch is however simply connected (in fact topologically equivalent to $R$), with trivial homotopy groups. The two zero divisors described above correspond to the asymptotes of the pair of hyperbolae. 
In the correspondence that we are trying to establish, the vacuum manifold in the conventional $U(1)$ field theory is defined by a circle, and as we shall see, for a theory with a hyperbolic symmetry the vacuum will have correspondingly the geometry of a hyperbola with two branches.
Along the same correspondence, the $U(1)$ modulus is invariant under circular rotation defined by the phase $e^{i\theta}$, and in the present case the modulus (\ref{modulus}) is invariant under the hyperbolic rotation $z\rightarrow e^{j\chi}z$, with $e^{j\chi}=\cosh \chi + j\sinh \chi$, and $\chi \in R$, and hence a noncompact parameter. Thus, hyperbolic rotations correspond to the connected component
of the Lie group $SO(1,1)$ that contains the unit group $I$. Additionally the modulus (\ref{modulus}) remains invariant under the ${\cal PT}$-like transformation
\begin{eqnarray}
(x,y)\rightarrow (-x,-y),
\label{pt}
\end{eqnarray}
and hence we can add the element $-I$, and we have the full $SO(1,1)$ group as the fundamental symmetry of the quadratic form (\ref{modulus}).
More details and discussions on the hyperbolic numbers and their applications in physics can be found  for example in \cite{ulrich},  \cite{pood}, and \cite{U1}; for our purposes we need only the basic aspects described above.

\section{Unifying inflaton-phantom cosmology with hyperbolic symmetries}
\label{uni}
Once we have introduced the appropriate complex unit, {\it unification} is actually a trivial issue; we just need the canonical and universal form for the kinetic term of a complex field $\psi=\phi_{1}+j\phi_{2}$ (instead of the aforementioned non-canonical term):
\begin{eqnarray}
{\cal L}_{K}= \partial_{\mu}\psi\partial^{\mu}\overline{\psi}=
\partial_{\mu}\phi_{1}\partial^{\mu}\phi_{1}-\partial_{\mu}\phi_{2}\partial^{\mu}\phi_{2};
\label{can}
\end{eqnarray}
which reproduces exactly the right hand side of Eq. (\ref{non}) and is, by construction, invariant under global hyperbolic
rotations; the generalization to local rotations is straightforward, and thus the hyperbolic electrodynamics, with hyperbolic counterparts of gauge fields, Higgs fields, and Nambu-Goldstone bosons emerges naturally within this approach (see \cite{Oscar} for more details); again, the field theoretic universal prescriptions work, with the Higgs mechanism included.
Note that the parallelism with the conventional $U(1)$ field theory is fulfilled; the internal symmetry in field space in the $U(1)$ case is defined completely by the complex unit $i$, and for the $SO(1,1)$ case by the hyperbolic unit $j$; any additional structure is not required.

Since our approach has been constructed by choosing an appropriate complex plane, therefore, the use of non-canonical forms in the action is neither technically nor conceptually needed in this respect. Indeed, {\it Euclideanization} through a field ``Wick rotation" has been avoided when attaining unification of scalar fields since it is, from the very beginning, unnatural. Moreover, on general grounds, the field theory {\it Euclideanization} is in its turn problematic in various senses, since an analytic continuation (with the conventional complex unit) and certain hidden structures  are involved in it (see \cite{penrose} for more details).  Although in the present context it is not required, it is important to clarify how a ``Wick rotation" will work in the hyperbolic plane. The  analytic con\-ti\-nua\-tion with $y\rightarrow j y$, in the quadratic form (\ref{modulus}) does not imply a change of sign and, thus, it does not correspond to an {\it Euclideanization}. More explorations along these lines are necessary, and they will be the subject of future communications.

On the other hand, non-canonical terms, such as those introduced in Eq. (\ref{non}), and those of the form $\partial \phi \partial\phi+c.c.$, $\phi^2+c.c.$, etc., are not invariant under circular or hyperbolic rotations, and in general are depending on the phase of the rotation. They can be used (without {\it Euclideanization}) as part of a mechanism that explicitly breaks a global symmetry, according to the ideas outlined in the introduction.
We shall consider this issue in section \ref{break}, by using a symmetry breaking term of the form $\phi^2+c.c.$

Now, by using our fundamental $SO(1,1)$-invariant expression (\ref{modulus}), we shall construct quartic-order potentials that, being bounded from below, allow us to study the realizations of spontaneous symmetry breaking along well-known universal prescriptions.

\section{Hyperbolic Mexican hat potential, and spontaneous symmetry breaking}
\label{mexican}
\begin{figure}[H]
  \begin{center}
    \includegraphics[width=.55\textwidth]{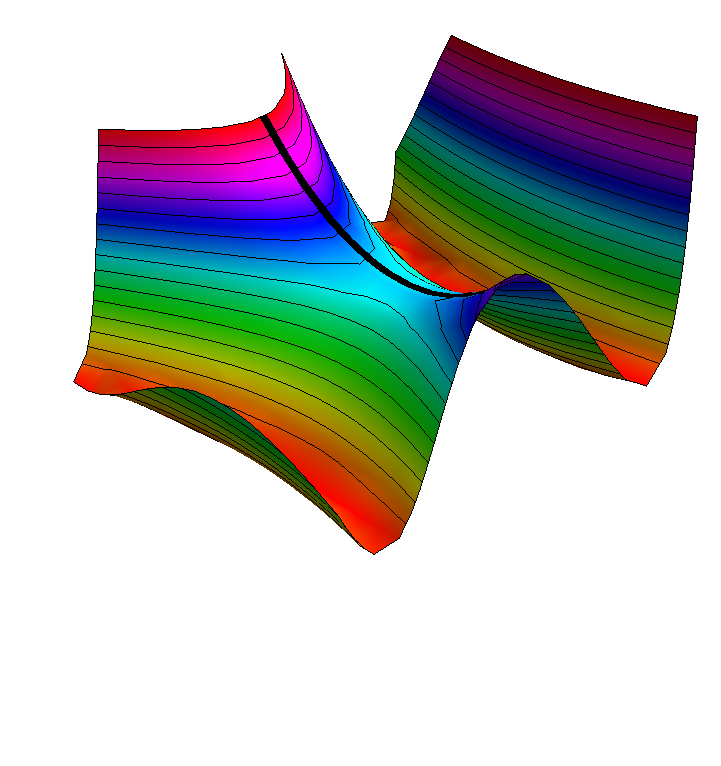}
  \caption{The potential (\ref{pot1}) for $m^2>0$, and $\lambda>0$; the black curve on the ridge represents the critical bifurcation points; the stable states are localized at the red
  regions, the hyperbola described by Eq. (\ref{vertices}). The scheme at hand shows a natural connection between the algebraic structure of the new complex plane and its underlying hyperbolic geometry, which is induced as the geometry (and topology) of the vacuum manifold, as opposed to the formulation used in \cite{Lorenz}.}  
  \label{infla1}
\end{center}
\end{figure}

Now we invoke the universal form of a potential containing quadratic mass, and quartic interaction terms, in a fully similar way to the $U(1)$ field theory,
\begin{equation}
     V(\psi,\overline{\psi}) = m^{2}\psi\overline{\psi} + \frac{\lambda}{6}\left(\psi\overline{\psi}\right)^{2} = m^{2}(v^{2}-w^{2}) + \frac{\lambda}{6} (v^{2}-w^{2})^{2}, \quad \psi = v+jw;
     \label{pot1}
\end{equation}
the degenerate vacuum corresponds to the hyperbola
\begin{equation}
     w^{2}_{0}-v^{2}_{0} = \frac{3}{\lambda}m^{2}, \quad   V(v_{0},w_{0}) = -\frac{3}{2} \frac{m^{4}}{\lambda},\quad  <w^{2}_{0}> = \frac{3}{\lambda}m^{2};
     \label{vertices}
\end{equation}
where $\lambda$ is a positive definite constant that guarantees the potential to be bounded from below and we have displayed the energy of the vacuum, and the v.e.v. of the field $w$  just at the vertices $(v_{0}=0, \pm w_0)$ of the hyperbola, around which the spontaneous symmetry breaking will be realized; these vertices are the only vacuum points where only one field develops a nonvanishing v.e.v.
Additionally such points will be relevant in the appearance of the new vacuum once the hyperbolic rotations are explicitly broken (see Figure \ref{infla2} in the next section).
Note that the hyperbolic symmetry of the potential (\ref{pot1}) is also present in the vacuum constraint (\ref{vertices})
through the invariance under the rotation $(v_{0}+j w_{0})\rightarrow e^{j\theta}(v_{0}+j w_{0})$, which is in fact responsible for its degeneracy; thus, the hyperbolic rotations preserve hyperbolas in the plane. This potential, that has appeared for the first time in \cite{Oscar},
is shown in Figure \ref{infla1}.

Hence, if the spontaneous symmetry breaking is performed by expanding the theory around the vertices (\ref{vertices}), then the diagonal mass matrix reads
\begin{eqnarray}
\left( \begin{array}{cc}
                 0 & 0 \\
                 0 & 2m^2w^2
                 \end{array} \right),
     \label{mm}
\end{eqnarray}  
which corresponds to a massive field $w$, and a massless field $v$ \cite{Oscar}; hence, the field $w$ that develops a nontrivial v.e.v. 
corresponds to the mode whose oscillations are ``transversal" to the hyperbolic valley.
Furthermore, since the mass term of the field $v$
has disappeared, it is understood as a Nambu-Goldstone boson
and corresponds to the mode whose oscillations lie in the hyperbolic valley. As we shall see in section \ref{break}, this Goldstone boson of the hyperbolic symmetry will acquire mass when that continuous global symmetry is explicitly broken, and then it becomes a pseudo-Goldstone boson.

\section{Breaking down the hyperbolic rotations}
\label{break}
\begin{figure}[H]
  \begin{center}
   \includegraphics[width=.65\textwidth]{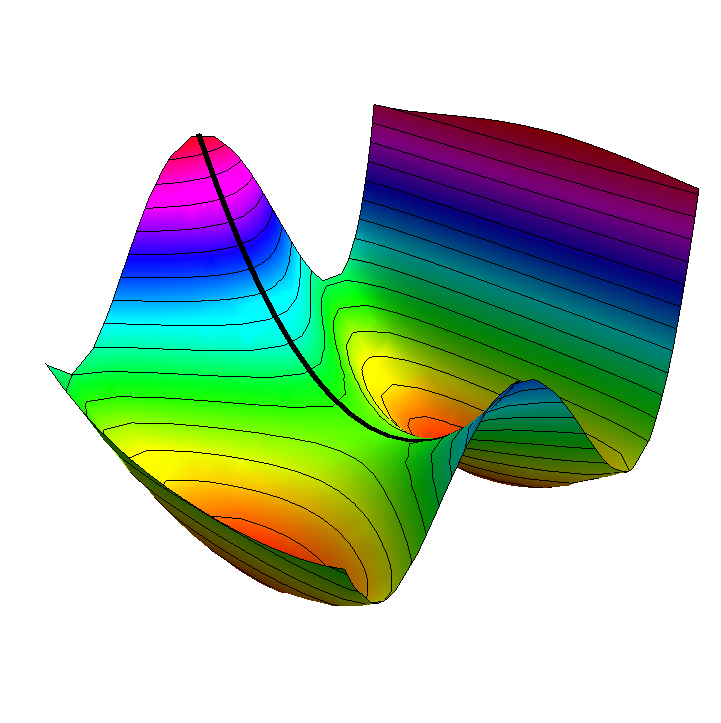}
  \caption{The potential (\ref{change1}); the vertices of the hyperbolic valley have been {\it depressed}  into the
   {\it bottom} of the red
  regions, where stable states are localized; the black curve on the ridge represents again the critical bifurcation points.}  
  \label{infla2}
\end{center}
\end{figure}

Following \cite{Riotto}, the global hyperbolic rotations of the potential   
 (\ref{pot1})
can be explicitly broken by adding a term depending on the hyperbolic phase,
\begin{equation}
    \Delta V(\psi,\overline{\psi}) = M^{2}_{p} (\psi^{2}+\overline{\psi}^{2}) = 2M^{2}_{p} (v^{2}+w^{2}),
    \label{change}
\end{equation}
 explicitly, if $z=\rho e^{j\chi}$ is the polar form for $z=x+jy$, then $x^2+y^2=\rho^{2}(2\cosh^{2}\chi-1)$;
  higher order terms of the form $\phi^{d}+\phi^{*d}$ can be considered, and they will be explored in forthcoming works. We
 have to realize that the expression (\ref{change}) is constructed on the ring with $z=x+jy$, on which the quadratic form $x^{2}+y^{2}$ is not invariant under hyperbolic rotations.
In any way, if one insists in the fact that the quadratic form $v^{2}+w^{2}$ is invariant under circular rotations as usual, the sum $V+\Delta V$ is not invariant under hyperbolic and/or circular rotations \footnote{Generalized commutative rings allow us to define objects that are invariant under both, circular and hyperbolic rotations; see concluding remarks.}. Explicitly, we have
\begin{equation}
     V+\Delta V = (m^{2}+2M^{2}_{p})v^{2} - (m^{2}-2M^{2}_{p})w^{2} + \frac{\lambda}{6} (v^{2}-w^{2})^{2};
     \label{change1}
\end{equation}
since the masses in the potential (\ref{change1}) are necessarily different from each other, the hyperbolic symmetry is effectively broken at the level of the quadratic mass terms; note however that the quartic $\lambda$-term is maintained invariant under hyperbolic rotations.
 On the basis of the following analysis, this modified potential is shown in Figure \ref{infla2} for certain choice of the parameters.
The critical points are
\begin{eqnarray}
     \! & & \! (v_{0}=0,\quad w_{0}=0), \quad {\rm Hess} = -4(m^{2}+2M^{2}_{p})(m^{2}-2M^{2}_{p}); \label{cp1} \\
     \! & & \! \big( v_{0}=0, \quad w^{2}_{0}=\frac{3}{\lambda}(m^{2}-2M^{2}_{p})\big), \quad {\rm Hess}=32M^{2}_{p}(m^{2}-2M^{2}_{p}); \label{cp2} \\
     \! & & \! \big( v^{2}_{0}=-\frac{3}{\lambda}(m^{2}+2M^{2}_{p}),\quad w_{0}=0\big) , \quad {\rm Hess}=-32M^{2}_{p}(m^{2}+2M^{2}_{p}); \label{cp3}
\end{eqnarray}
where the corresponding Hessian is displayed; note that the hyperbola (\ref{vertices}) has disappeared as degenerate vacuum. Therefore, a potential bounded from below can be obtained by restricting all parameters, $M^{2}_{p}$, $m^{2}+2M^{2}_{p}$, $m^{2}-2M^{2}_{p}$, and $\lambda$, to be positive; the zero-energy point (\ref{cp1}) is a saddle point with $Hess(0,0)<0$, and the points given by (\ref{cp2}) correspond to the only (global) minimum points with $Hess(v_{0}=0,\pm w_{0})>0$, and they are localized at the potential {\it bottom}, in the two red regions.
Note that the points described by (\ref{cp3}) are not realizable as critical points since $v^{2}_{0}<0$ for this choice of parameters.

Therefore, the original degenerate vacuum, defined by the hyperbolic valley in Figure \ref{infla1}, has been hollowed out in the two vertices, which have now the following adjusted energy and v.e.v.
\begin{eqnarray}
     (V+\Delta V)_{0} \! & = & \! -\frac{3}{2\lambda}(m^{2}-2M^{2}_{p})^{2},\quad  \quad  <w^{2}_{0}>_{[V+\Delta V]} =\frac{3}{\lambda}(m^{2}-2M^{2}_{p});
\label{new1} \\
 V _{0} \! & = & \! -\frac{3}{2\lambda}m^{4}, \qquad  \quad <w^{2}_{0}>_{[V]}=\frac{3}{\lambda}m^{2}; \label{new2}
\end{eqnarray}
where we have displayed for direct comparison, the energy and v.e.v. described in the previous section for the potential $V$; therefore,
the following inequalities are satisfied, $V+\Delta V > V$, and $ <w^{2}_{0}>_{[V+\Delta V] }\quad < \quad <w^{2}_{0}>_{[V] }$.

Geometrically the original vacuum manifold, the hyperbolic valley, has been hollowed out at the vertices; from the topological point of view, the two copies of $R$ representing the two branches of the hyperbola, have undergone a retraction to a pair of points. Such a retraction will constitute a topological argument for the stability  of certain type of topological defects that will be considered in Section \ref{persistence}.

\subsection{Spontaneous symmetry breaking around the new vacuum}
\label{vacua}
The expansion around the vacuum point $\psi_{0}=(0,w_{0})$ given in Eq. (\ref{new1}), with $\psi+\psi_{0}=v+j(w+w_{0})$, leads to
\begin{eqnarray}
(V+\Delta V)(\psi+\psi_{0})=   \Big(\left[m^2+2M^2_{p} \right]-\left[m^2-2M^2_{p} \right]
\Big)v^2+ 2(m^2-2M^2_{p})w^2+ {\rm higher-terms};
\end{eqnarray}
 then the diagonal mass matrix 
reads
\begin{eqnarray}
\left( \begin{array}{cc}
                 4M^2_{p}v^2 & 0 \\
                 0 & 2(m^2-2M^2_{p})w^2
                 \end{array} \right),
     \label{mm2}
\end{eqnarray}
which reduces consistently to the mass matrix (\ref{mm}) in the limit $M_{p}\rightarrow 0$; thus, the Goldstone boson of the hyperbolic symmetry has acquired an effective mass, which is just that mass lost by the field $w$. As it is well known, this is exactly the case arising in QCD, where the pion corresponds to a pseudo-Goldstone boson of the chiral symmetry \cite{Riotto}.

The appearance of two massive modes in Eq. (\ref{mm2}) can be understood geometrically from Figure \ref{infla2}; first the hyperbolic valley no longer exists, and hence there are no modes of zero frequency, and both oscillations are now around the new vacuum localized at the potential {\it bottom}.

\section{Quartic deformation}
\label{qua}
\begin{figure}[H]
  \begin{center}
   \includegraphics[width=.75\textwidth]{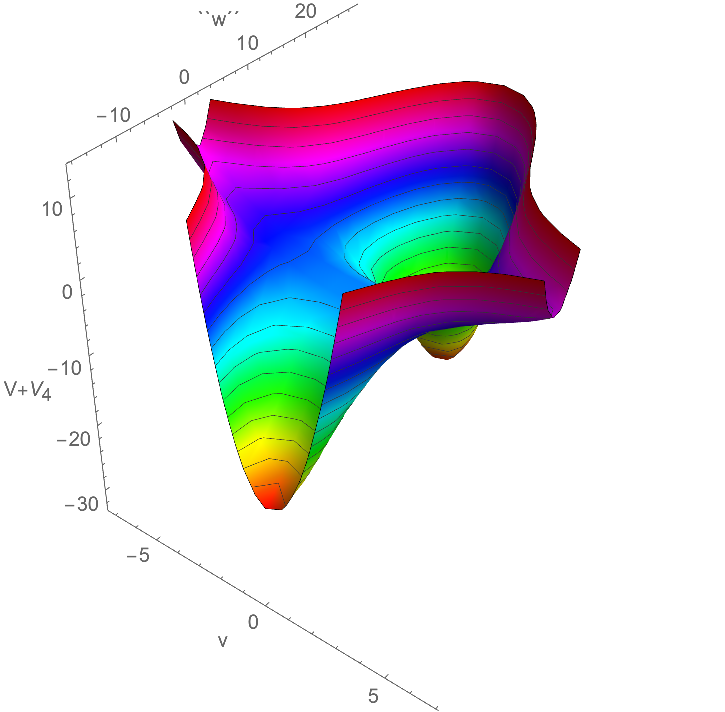}
  \caption{The potential (\ref{quartic1});  the
   {\it bottom} of the red
  regions corresponds again to global minima;  the critical bifurcation points at the ridge lie now, as opposed to those points in the figure \ref{infla2}, on a {\it parabolic valley}.}  
  \label{quarticfig}
\end{center}
\end{figure}

Similarly to the quadratic deformation (\ref{change}), we study now a quartic deformation of the form
\begin{equation}
     \Delta V_{4} (\psi,\bar{\psi}) = \lambda_{p} (\psi^{4} + \bar{\psi}^{4}) = 2 \lambda_{p} (v^{4} + 6v^{2}w^{2} + w^{4}),
     \label{quartic}
\end{equation}
and thus, the effective potential reads
\begin{equation}
     V + \Delta V_{4} = m^{2} (v^{2}-w^{2}) + (12\lambda_{p} - \frac{1}{3}\lambda) v^{2}w^{2} + (2\lambda_{p} + \frac{\lambda}{6}) (v^{4}+w^{4});
     \label{quartic1}
\end{equation}
note that such a deformed potential retains the discrete symmetries $(v.w)\rightarrow (-v,-w)$, $(v,w)\rightarrow (-v,w)$, and $(v,w)\rightarrow (v,-w)$ (see figure \ref{quarticfig}). Critical points correspond to
\begin{eqnarray}
    \! & & \! (v_{0}=0, \quad w_{0}=0), \quad Hess = -4m^{4}; \label{quartic2} \\
    \! & & \! (v_{0}=0, \quad <w_{0}>^{2}=3 \frac{m^{2}}{\lambda +12\lambda_{p}}), \quad Hess = 8(48)m^{4}\frac{\lambda_{p}}{\lambda +12\lambda_{p}}; \label{quartic3} \\
    \! & & \! (<v_{0}>^{2}= -\frac{3m^{2}}{\lambda +12\lambda_{p}}, \quad w_{0}=0), \quad Hess =12(32)m^{4}\frac{\lambda_{p}}{\lambda +12\lambda_{p}}; \label{quartic4} 
\end{eqnarray}
there no exist critical points with $(v_{0}\neq 0, w_{0}\neq 0)$, and with $\lambda_{p}\neq 0$; furthermore, the $(0,0)$ critical point (\ref{quartic2}) is always a saddle point, indistinct to the sign of $m^{2}$. The realization of (\ref{quartic3}) as critical points requires simultaneously that
\begin{equation}
     m^{2}>0, \quad \lambda +12\lambda_{p} >0; \label{quartic5}
\end{equation}
note that under such constraints, the critical points (\ref{quartic4}) are not realizable.

The simplest realization of the second constraint in (\ref{quartic5}) is with
\begin{eqnarray}
     \lambda >0, \quad \lambda_{p}>0; \label{quartic6}\\
     \quad \frac{\partial}{\partial v^{2}} (V+\Delta V_{4})=2(48)\frac{m^{2}\lambda_{p}}{\lambda +12\lambda_{p}}>0, \quad \frac{\partial}{\partial w^{2}}(V+\Delta V_{4})=4m^{2}>0,    \quad Hess (v_{0}=0, \quad w_{0}\neq 0)>0; \label{quartic7}
\end{eqnarray}
therefore, we have obtained a potential with two global minima (and thus bounded from below), and the $(0,0)$ as a saddle point; the profile of this potential is basically that described previously in figure (\ref{infla2}). However, there exists another possibility, namely, the second constraint in (\ref{quartic5}) is realizable also by retaining the  constraint $\lambda_{p}>0$ (see the Hess in Eq. (\ref{quartic3}), and the second derivatives in Eq. (\ref{quartic7})), and $\lambda<0$, and with the constraint
\begin{equation}
     12\lambda_{p}>|\lambda|; \qquad (\lambda_{p}>0, \quad \lambda<0);
               \label{quartic8}
\end{equation}
thus, the restrictions (\ref{quartic7}) are also satisfied, and we obtain again a potential bounded from below,
which
is illustrated in the figure (\ref{quarticfig}).
Note that, in similarity to the potential described in the figure \ref{infla2},  the hyperbolic valley has undergone also a retraction to a pair of points.  Along the same lines of subsection \ref{vacua}, the expansion around the vacuum point $\psi_{0}=(0,w_{0})$ given in the Eq. (\ref{quartic3}), the mass matrix takes the form
\begin{eqnarray}
\left( \begin{array}{cc}
                 16<w_{0}>^2v^2 & 0 \\
                 0 &2m^2w^2
                 \end{array} \right),
     \label{mm2}
\end{eqnarray}
which reduces consistently to the mass matrix (\ref{mm}) in the limit $\lambda_{p}\rightarrow 0$; note that both masses in this matrix are positive for the two cases described in Eqs. (\ref{quartic6}), and (\ref{quartic8}).

We would like to notice that even though the potential (\ref{infla1}) possesses some of the properties of the Hybrid Natural Inflation's potential \cite{Linde}, the potentials described in the figures (\ref{infla2}) and (\ref{quarticfig})  with the broken hyperbolic symmetry do not yield the Hybrid Inflation scenario because the inflaton field does not change its sign just before ending inflation, i.e. before it triggers the waterfall field that will attain global minimum. However, a different $\Delta V$ could give an adequate description of the Hybrid Inflation scenario or one of its suitable modifications within the framework of Cosmology. This is a task which is under current research at the moment.\footnote{A.H.-A. thanks Gabriel Germ\'an for useful discussions on this issue.}

\section{On the formation of topological defects}
\label{topo}
\subsection{Topology of vacuum and topological defects}
The topology of the vacuum manifold determines the type of topological defects that may be formed through the Kibble mechanism; in the case with unbroken hyperbolic symmetry discussed in section \ref{mexican}, the two branches of the hyperbolic valley are topologically equivalent to two copies of $R^1$, which in their turn are homotopically equivalent to two points, and thus there are no topological defects associated with continuous symmetries. No monopoles, no strings, no textures, etc. are admitted; however, the fact that two disconnected copies of $R^1$ constitute the vacuum leads to a nontrivial $0$-homotopy group, and hence {\it domain walls} are possible \cite{Oscar}. Furthermore, as we have seen in section \ref{break}, the first scenario for the possible topological defects that the theory admits is preserved under an explicit breakdown of the hyperbolic rotations, since the hyperbolic valley will be hollowed out in the two vertices; an exactly similar conclusion is valid for the potential described in section \ref{qua}.

Therefore, the breakdown of the global hyperbolic symmetry through a quadratic deformation (section \ref{break}), or a quartic deformation (section \ref{qua}), has reduced the original hyperbolic valley to (homotopically equivalent) two points, and the {\it domain walls} are maintained as the only topological defects that may be formed through the Kibble mechanism; since the vacua consist of two points $\pm w_{0}$, two regions in the physical space will form with $<w>=\pm w_{0}$, which are separated by a region of false vacuum with $<w>=0$, which corresponds to the $(0,0)$ saddle point.

\subsection{Persistence of Domain walls and constraints on cosmic strings}
\label{persistence}
With the motivation that stable topological defects of scalar fields can contribute to dark energy/matter of the universe,  the possibility of detecting a network of domain walls in the galactic environment has been considered recently \cite{wall}; with a global network of synchronized optic magnetometers the authors considered the viability of detecting such defects, even when the gravitational and astrophysical constraints are taken into account.

In the present formulation, the deformation $\Delta V$ constructed by the addition of quadratic mass terms, does not eliminate {\it the domain-wall problem}; however, in the light of the search of the domain walls discussed above, the persistence of such defects can be interpreted
in a converse sense: the formation of domain walls is stable under that explicit breaking of the global hyperbolic symmetry; this result is supported by a topological argument: the retraction of the hyperbolic valley to two points homotopically equivalent, and described in the figures \ref{infla2}, and \ref{quarticfig}.

Nowadays there is no experimental evidence for the existence of topological defects at cosmological scale; for example, new severe constraints on cosmic strings parameters were established recently by analyzing data collected by the LIGO and Virgo gravitational wave detectors \cite{ligo}.
 From the theoretical point of view
it is interesting that within the present scheme, most cosmological defects have been ruled out, due to the presence of the hyperbolic symmetry, retaining only those defects associated with discrete symmetries.  

\section{ Action with hyperbolic symmetry, non-linear diffusion, and quantum mechanics} 
\label{hamilton}
Now we will consider the non-relativistic limit of the complete actions, the kinetic term discussed in section \ref{uni}, plus the potentials constructed in the previous sections, and its connection with nonlinear diffusion systems, and quantum mechanics;
in a d-dimensional background the action with global hyperbolic symmetry  reads
\begin{equation}
     S[\varphi ,\psi] = \int dx^{d} \left[\partial^{\mu} \psi\partial_{\mu}\bar{\psi} - m^{2}\psi\bar{\psi} - \frac{\lambda}{6} (\psi\bar{\psi})^{2}\right],
     \label{alpha}
\end{equation}
where $m^{2}>0$, and the self-interacting constant $\lambda$ is  assumed to be positive in order to have the energy bounded from below;
 the equation of motion  obtained from the action (\ref{alpha}) with respect  to the field  $\bar{\psi}$   can be written  in the following form after introducing  convenient units, and assuming a space-time decomposition, 
\begin{equation}
     -\hbar^{2} \frac{\partial^{2}}{\partial t^{2}}\psi + \hbar^{2}c^{2}\partial^{i}\partial_{i}\psi + m^{2}c^{4}\psi + \frac{\lambda mc^{2}}{3}(\bar{\psi}\psi)\psi= 0; \label{gamma1}
\end{equation}
the non-relativistic limit can be obtained by considering first the field transformation
  $\psi\rightarrow\phi(x,t) e^{jmc^{2}t/\hbar}$ \cite{greiner}, and thus the above equation reduces to
 \begin{equation}
     -\hbar^{2} \frac{\partial^{2}\phi}{\partial t^{2}} + j2mc^{2}\hbar \frac{\partial\phi}{\partial t} + \hbar^{2}c^{2}\nabla^{2}\phi + \frac{\lambda mc^{2}}{3} (\bar{\phi}\phi)\phi =0,
\end{equation}
because of the units that we have used, we can observe that the first term lacks a $c^{2}$ factor and can be neglected respect to the other ones, obtaining as the non-relativistic limit the following non-linear Schr\"odinger-like wave equation
\begin{equation}
     j\hbar \frac{\partial\phi}{\partial t} = - \frac{\hbar^{2}}{2m} \nabla^{2}\phi - \frac{\lambda}{6}(\bar{\phi}\phi)\phi, \label{gamma2}
\end{equation}
where $\phi$ is a hyperbolic complex field $\phi=\phi_{1}+j\phi_{2}$;
this equation represents then the hyperbolic version of the usual non-linear Schr\"odinger equation based in the usual complex unit $i$; the correspondence between these equations will be explored in future works along the lines developed in \cite{kocik}, where the duality between diffusion systems (formulated on the hyperbolic plane) and quantum mechanics (formulated on the usual complex plane) was established.

\section{Concluding remarks} 
\subsection{Comparison with inflationary models}
\label{comparison1}
On the other hand, in \cite{hyb} the following hybrid inflation potential\footnote{Hybrid inflationary models encompass properties of chaotic inflation with $V(\phi)=\frac{m^2\phi^2}{2}$ and the usual theory with a spontaneous symmetry breaking mechanism described by the potential $V(\sigma)=\frac{1}{4\lambda}\left(M^2-\lambda\sigma^2\right)$.} was proposed {\it ad hoc} to address the problem of the preheating phenomenon, without
invoking any symmetry principle,
\begin{equation}
     V(\sigma, v) = \frac{\lambda}{4}(\sigma^{2}-\sigma_{0}^{2})^{2} + \frac{1}{2}g^{2}\phi^{2}\sigma^{2}, \label{hybri1}
\end{equation}
where $\sigma$ is a Higgs field, which remains a physical degree of freedom after the Higgs effect in an underlying gauge theory with spontaneous symmetry breaking, $\sigma_{0}$ a critical point and $\phi$ is the inflaton field; this potential is illustrated in the Figure 1 of \cite{hyb}, however, the profile of such a potential is basically the ones described in Figure \ref{infla2}, and \ref{quarticfig}, within the scheme at hand: a set of critical bifurcation points on the ridge, and two points in the valley as stable minima.    
Therefore, on the basis of the unification of scalar fields through a hyperbolic complex field and the use of symmetry principles, we have been able to generate the necessary ingredients for an inflationary scenario.

 It is worth mentioning as well that within the framework of the original hybrid inflationary model \cite{Linde}, which makes use of a pair of interacting scalar fields with an effective potential of the form (\ref{hybri1}); the inflaton scalar field is initially responsible for the slow-roll inflationary phase along one field direction, whereas inflation may end by a rapid rolling of the Higgs field along a transverse field direction. As soon as the inflaton field reaches its minimum at the last stage of inflation, the inflation is driven not by the energy density of the inflaton field $V(\phi)$, but by the noninflationary potential $V(\sigma)$: Immediately after the inflaton attains its critical value, the waterfall scalar field $\sigma$ is triggered by the inflaton $\phi$ and acquires a very rapid rolling, moving faster and faster (hence the name waterfall) towards the minima of the effective potential, ending inflation and providing a qualitatively suitable scenario for studying inflation and the way it ends in cosmology. It would be interesting to analyze in more detail the cosmological consequences that these kind of hybrid potentials carry for inflationary models within the framework of the above mentioned hyperbolic field theory. This is an issue that is currently under research.

\subsection{On the correspondence between unitary and orthogonal symmetries and the generalized hyperbolic ring}
\label{cr}
The complex formalism used in this scheme consists basically in the substitution of the usual elliptic complex unit $i$ by the hyperbolic unit $j$; however, a more general commutative ring can be generated by the incorporation of the new complex unit $j$  to the usual one $i$; such a formalism allows us to unify compact and noncompact symmetry groups. For example, hypercomplex electrodynamics emerges as a generalization of the hyperbolic electrodynamics commented previously \cite{Oscar}. Furthermore, within this generalized ring a hypercomplex scalar field encoding four real fields can be defined; thus, a field theory with an internal symmetry that unifies the $U(1)$ compact group with the $SO(1,1)$ noncompact group can be constructed \cite{alex}. The algebraic structure
of the ring allows us to avoid normal ordering of operators for controlling the vacuum energy, which reduces to an exact zero value;
an alternative approach based on a bicomplex algebra is developed in \cite{max}.
These results show the advantages of using alternative and complementary complex units, and the necessity  of going  beyond the traditional use of the conventional complex unit.

\begin{table}[htb]
\centering
\caption{The correspondence $U(1)\leftrightarrow SO(1,1)$, through an universal functional dependence in the pair $(\psi,\psi^{*})$;
``dof" means degrees of freedom, for the exact symmetry we have $\Delta V=0$, whereas for the broken symmetry we get $\Delta V\neq0$. In the complex units column, we indicate the algebraic structure associated with each complex unit. For $\Delta V=0$ the conserved current has the universal functional form $J^{\mu} = \iota[(\partial^{\mu}\bar{\psi})\psi - (\partial^{\mu}\psi)\bar{\psi}]$, with $\psi=\psi_{1}+\iota\psi_2$, and thus $\iota=i$ for the $U(1)$ field theory, and $\iota=j$ for the $SO(1,1)$ field theory .}
\label{mylabel}
\begin{tabular}{l|c|c|c|c|c|c|c|}
\cline{2-8}
 & \multicolumn{7}{c|}{{$\partial_{\mu}$$\psi$$\partial^{\mu}$${\psi}^{*}$+$m^{2}$$\psi$${\psi}^{*}$+$\lambda$($\psi$${\psi}^{*})^{2}$+$\Delta$V}} \\ \hline
\multicolumn{1}{|l|}{\multirow{2}{*}{}} & \multirow{2}{*}{\begin{tabular}[c]{@{}c@{}}complex\\ unit\end{tabular}} & \multicolumn{2}{c|}{local dof} & \multicolumn{2}{c|}{topological defects} & \multicolumn{2}{c|}{\begin{tabular}[c]{@{}c@{}}homotopy \\ constraint\end{tabular}} \\ \cline{3-8} 
\multicolumn{1}{|l|}{} &  & exact & broken & exact & broken & exact & broken \\ \hline
\multicolumn{1}{|c|}{\begin{tabular}[c]{@{}c@{}}compact\\ U (1)\end{tabular}} & i/field & \multicolumn{2}{c|}{2} & strings & \begin{tabular}[c]{@{}c@{}}domain\\ walls\end{tabular} & $\pi_{1} \neq 0$ & \begin{tabular}[c]{@{}c@{}}$\pi_{0} = 2$, \\ $\pi_{n} = 0$, $n\ge{1}$\end{tabular} \\ \hline
\multicolumn{1}{|c|}{\begin{tabular}[c]{@{}c@{}}noncompact\\ SO (1,1)\end{tabular}} & j/ring & \multicolumn{2}{c|}{2}  & \multicolumn{2}{c|}{\begin{tabular}[c]{@{}c@{}}domain\\ walls\end{tabular}}  & \multicolumn{2}{c|}{\begin{tabular}[c]{@{}c@{}}$\pi_{0} = 2$, \\ $\pi_{n} = 0$, $n\ge{1}$\end{tabular}} \\ \hline
\end{tabular}
\end{table}

\newpage

{\bf Acknowledgements}
This work was supported by the Sistema Nacional de Investigadores (SNI, Mexico), and the Vicerrector\'ia de Investigaci\'on y Estudios de Posgrado (VIEP-BUAP). Graphics have been made using Mathematica.

\end{document}